\documentclass[conference]{IEEEtran}
\IEEEoverridecommandlockouts
\usepackage{cite}
\usepackage{amsmath,amssymb,amsfonts}
\usepackage{algorithmic}
\usepackage{graphicx}
\usepackage{subcaption}
\usepackage{multirow}
\usepackage{textcomp}
\usepackage{array}
\usepackage{float}
\usepackage{url}
\usepackage{dblfloatfix}
\usepackage[usenames, dvipsnames]{color}
\usepackage[font=small,labelfont=small]{caption}
\def\BibTeX{{\rm B\kern-.05em{\sc i\kern-.025em b}\kern-.08em
    T\kern-.1667em\lower.7ex\hbox{E}\kern-.125emX}}
\begin{document}

\title{A Neural Model for Generating Natural\\Language Summaries of Program Subroutines
%\title{Generating Abstractive Summaries of Java Methods 
\thanks{\vspace{-0.08cm}This work is supported in part by \emph{  NSF  CCF-1452959,
CCF-1717607, and CNS-1510329 grants}.}
}
\author{\IEEEauthorblockN{Alexander LeClair\IEEEauthorrefmark{1}, Siyuan Jiang\IEEEauthorrefmark{2}, Collin McMillan\IEEEauthorrefmark{1}}
\IEEEauthorblockA{\IEEEauthorrefmark{1}\textit{Dept. of Computer Science and Engineering} \\
	\textit{University of Notre Dame}\\
	Notre Dame, IN, USA \\
	Email: \{aleclair, cmc\}@nd.edu
}
\IEEEauthorblockA{\IEEEauthorrefmark{2}\textit{Dept. of Computer Science}\\
    \textit{Eastern Michigan University}\\
    Ypsilanti, MI, USA\\
    Email: sjiang1@emich.edu
}
}
%\author{\IEEEauthorblockN{\emph{Author list redacted for blind review.}}
%	\IEEEauthorblockA{\textit{~} \\
%		\textit{~}\\
%		~ \\
%		~
%}}

\maketitle

\begin{abstract}
Source code summarization -- creating natural language descriptions of source code behavior -- is a rapidly-growing research topic with applications to automatic documentation generation, program comprehension, and software maintenance.  Traditional techniques relied on heuristics and templates built manually by human experts.  Recently, data-driven approaches based on neural machine translation have largely overtaken template-based systems.  But nearly all of these techniques rely almost entirely on programs having good internal documentation; without clear identifier names, the models fail to create good summaries.  In this paper, we present a neural model that combines words from code with code structure from an AST.  Unlike previous approaches, our model processes each data source as a separate input, which allows the model to learn code structure independent of the text in code.  This process helps our approach provide coherent summaries in many cases even when zero internal documentation is provided.  We evaluate our technique with a dataset we created from 2.1m Java methods.  We find improvement over two baseline techniques from SE literature and one from NLP literature.
\end{abstract}

\begin{IEEEkeywords}
automatic documentation generation, source code summarization, code comment generation
%\vspace{-0.14cm}
\end{IEEEkeywords}

%\vspace{-0.08cm}
\section{Introduction}
A ``summary'' of source code is a brief natural language description of that section of source code~\cite{mcburney2016automatic}.  One of the most common targets for summarization are the subroutines in a program; for example, the one-sentence descriptions of Java methods widely used in automatically-formatted documentation e.g. JavaDocs~\cite{kramer1999api}.  These descriptions are useful because they help programmers understand the role that the subroutine plays in a program -- empirical studies have repeatedly shown that understanding the role of the subroutines in a program is a crucial step to understanding the program's behavior overall~\cite{von1995program, letovsky1987cognitive, cornelissen2009systematic, maletic2001supporting}.  Even a short summary of a subroutine e.g. ``returns the player's hitpoint count'' can tell a programmer a lot about that subroutine and the program as a whole.  

A holy grail of software engineering research has long been to generate these summaries automatically.  Forward~\emph{et al.} pointed out in 2002 that ``software professionals value technologies that improve automation of the documentation process,'' and ``that documentation tools should seek to better extract knowledge from core resources'', such as source code~\cite{forward2002relevance}.  However, the state-of-the-practice has barely changed since that time for tool support for automated documentation generation.  Tools such as JavaDoc~\cite{kramer1999api} and Doxygen~\cite{DoxygenWebsite} automate the format and presentation of documentation, but still leave programmers with the most labor-intensive effort of writing the text and examples.

Research into generation of natural language descriptions of code has come to be known as ``source code summarization''~\cite{eddy2013evaluating}, with significant effort focused on generation of summaries of subroutines: For several years, significant progress was made based on content selection and sentence templates~\cite{hill2009automatically, haiduc2010use, sridhara2010towards, rastkar2011generating, mcburney2016automatic, moreno2013automatic} or even somewhat-idiosyncratic solutions such as mimicking human eye movements~\cite{rodeghero2015eye}.  However, as in many research areas and as chronicled in a recent survey by Allamanis~\emph{et al.}~\cite{allamanis2017survey}, these techniques have largely given way to AI based on big data input.

The inspiration for a vast majority of efforts into AI-based code summarization originates in neural machine translation (NMT) from the natural language processing research community.  An NMT system converts one natural language into another.  It is typically thought of in terms of sequence to sequence (seq2seq) learning, in which an e.g. English sentence is one sequence and is converted into a equivalent target sequence representing a e.g. French sentence.  In software engineering research, machine translation can be considered as a metaphor for source code summarization: the words and tokens in the body of a subroutine are one sequence, while the desired natural language summary is the target sequence.  This application of NMT to code summarization has shown strong benefits in a variety of applications~\cite{oda2015learning, allamanis2016convolutional, iyer2016summarizing, jiang2017automatically, yin2018mining, hu2018deep}.

However, an Achilles' heel to nearly all source code summarization techniques is a reliance on programmers having written high quality internal documentation in the form of identifier names or comments.  In order to generate a meaningful summary, meaningful words must be observed in the body of the subroutine.  In traditional NMT, this is accepted because a natural language input sentence will definitely have words related to the output target sentence.  But in software, the words in code are not actually related to the behavior of that code.  A subroutine's behavior is dictated by the structure of programming language keywords and tokens that define control flow, data flow, etc.  These differences between code and natural language are a barrier to improving performance in several AI applications to software engineering, as Hellendoorn~\emph{et al.}~\cite{hellendoorn2017deep} pointed out, to some controversy, at FSE'17.

In this paper, we present a neural model for summarizing subroutines.  Our model combines two types of information about source code: 1) a word representation treating code as text, and 2) an abstract syntax tree (AST) representation.  A distinguishing factor of our model compared to earlier approaches is that we treat both representations separately.  Previous techniques have shown promise by annotating a word representation with AST information~\cite{hu2018deep}, but ultimately the annotated representation is sent as a single sequence through a standard seq2seq model.  In contrast, our model accepts two inputs, one for the word representation and one for the AST.  The advantage is that we are able to treat each input with differently, which increases the flexibility of our approach, as we will show in this paper.

In essense, the neural model we propose involves two uni-directional gated recurrent unit (GRU) layers: one to process the words from source code, and one to process the AST.  We modify the SBT AST flattening procedure by Hu~\emph{et al.}~\cite{hu2018deep} to represent the AST.  We then use an attention mechanism to attend words in the output summary sentence to words in the code word representation, and a separate attention mechanism to attend the summary words to parts of the AST.  We concatenate the vectors from each attention mechanism to create a context vector.  Finally, we predict the summary one word at a time from the context vector, following what is typical in seq2seq models.

We evaluate our technique in two stages.  First, we collect over 51m Java methods from the Sourcerer repository~\cite{Lopes+Bajracharya+Ossher+Baldi:2010}, and preprocess them to form a dataset of around 2.1m methods with suitable JavaDoc summary comments.  We divide the dataset into training/validation/testing sets and perform a set of tests comparing results from our model to three competitive baselines.  We call this the \textbf{standard experiment}, because it conforms to common practice in both SE and NLP venues.

Second, to evaluate the limits of our model in a scenario without words from source code, we repeat the standard experiment using only the AST for each Java method -- in this study, we assume no code words are available, as in obfuscated code, poorly-written code, or situations in which there is only bytecode (from which an AST can be extracted but code words are likely to have been removed during compilation).  This ``no code words'' experiment simulates a situation unique to the SE domain and, as we will show, is far more difficult than the standard application of NMT in which a programmer provides useful keywords.  We call this the \textbf{challenge experiment}.

Our results, in a nutshell, are:
1) In the standard experiment, our model and the competitive NLP baseline provide comparable performance but with orthogonal predictions, implying that they are good candidates for ensemble decoding.  An ensemble provides state-of-the-art performance of 20.9 BLEU (an 8\% improvement over the nearest baseline).
2) In the challenge experiment, our model achieves 9.5 BLEU, versus 0 for any baseline.
%We demonstrate that it is possible to create summaries for many subroutines when given only the AST structure.
This is a significant step forward in source code summarization, since it requires zero meaningful code words.  We release all data, code, and implementation via our online appendix (see Section~\ref{sec:reproducibility}).

\section{Problem and Overview}
%\vspace{-0.1cm}
%\vspace{-.5cm}
We target the problem of source code summarization of subroutines -- automatic generation of natural language descriptions of subroutines.  Specifically, we target summarization of Java methods, with the objective of creating method summaries like those used in JavaDocs.  While we limit the scope of the experiments in this paper to a large Java dataset, in principle the techniques described in this paper are applicable to any programming language that has subroutines, from which an AST can be computed, and from which text e.g. identifier names can be extracted.  Our scoping of our target problem is consistent with the problem definition in many papers on code summarization~\cite{hu2018deep, mcburney2016automatic, rastkar2011generating, sridhara2010towards, richardson2017code2text}.

A solution to this problem would have many practical applications.  The primary practical application would be in automatic documentation generation, to help programmers write documentation more quickly, as well as understand code that has not been documented.  Of the 51m Java methods we found in the Sourcerer dataset, only about 10\% have any sort of method summary, and only about 4\% contain summaries that met basic quality filters we define in Section~\ref{sec:corpus}.  In our view, it seems likely that more than 4\% ``should'' be documented well, and an automatic summary generator would help improve the amount of code that could be documented.

But more generally, our goal for this paper is to also contribute to an ongoing academic debate about how to represent source code to solve software engineering problems using AI.  As mentioned, there is reasonable doubt~\cite{hellendoorn2017deep} that neural-based techniques are even appropriate for software engineering data; a recent workshop at AAAI'18~\cite{NL4SEAAAI:2018} focused heavily on this debate.  Given the long history of AI use to solve SE problems~\cite{xie2018intelligent}, our sincere hope for this paper is to provide insight into ways to build neural models of SE data, even for researchers outside of the specific task of code summarization.  We have made significant efforts to keep our data and techniques public and reproducible (see Section~\ref{sec:reproducibility}) to help these other researchers as much as possible.

An overview of this paper is below.  In the next section we cover background and related technologies.  Then, we introduce our proposed neural model.  We then describe how we obtained and processed the Java datasets we use.  We conduct the standard and challenge experiments on the same set of Java methods.  Finally, we spend significant space on examples and discussion.  We feel an in-depth look at examples where the model worked and did not will provide key insights for improving or adapting the model in the future.

\begin{figure}[!b]
	\centering
	\vspace{-0.5cm}
	\includegraphics[width=0.30\textwidth]{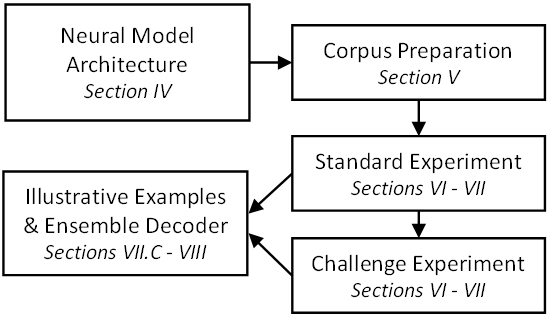}
	\vspace{-0.1cm}
	%\caption{An overview of this paper.}
	\label{fig:overview}
	%\vspace{-0.1cm}
\end{figure}

\section{Background and Related Work}
\label{sec:background}

This section covers the supporting technologies behind our work, plus related work in source code summarization.

%\subsection{Structure-based Traversal (SBT)}
\subsection{Source Code Summarization}

Related work in source code summarization can be broadly classified as either AI/data-driven or heuristic/template-driven.  

\subsubsection{Data-driven} Among data-driven techniques, recent work by Hu~\emph{et al.}~\cite{hu2018deep} is the most-closely related to this paper.  That work proposes to use the AST to annotate the words in the source code, and then use the annotated representation as input to a seq2seq neural model.  The model itself is an off-the-shelf encoder-decoder; the main advancement is the AST-annotated representation called Structure-based Traversal (SBT).  SBT is essentially a technique for flattening the AST and ensuring that words in the code are associated with their AST node type.  For example, the code {\small\texttt{request.remove(id)}} becomes:

\vspace{-0.2cm}
{\small
\begin{verbatim}
 ( MethodInvocation
 ( SimpleName_request ) SimpleName_request
 ( SimpleName_remove ) SimpleName_remove
 ( SimpleName_id ) SimpleName_id
 ) MethodInvocation
\end{verbatim}
}
\vspace{-0.2cm}

The intent is that the words ``request'', ``remove'', and ``id'' be associated with the context in which they appear.  In this case, a MethodInvocation node.  The SBT representation forms an important baseline for comparison in our experiments in later sections.  A casual reader will note that SBT was shown in that paper to obtain remarkable performance of 38 BLEU, but we caution that this is not directly comparable to the results in our experiments.  The reason is that in \cite{hu2018deep}, the dataset was split by function, so the training, validation, and test sets contain random selections of functions in the entire dataset.  In contrast, we split by \emph{project}.  In \cite{hu2018deep}, functions from the same project can be in both the training and test sets.  In our experiments, all methods from a project are either training, validation, or test.  In addition, we performed other preprocessing such as auto-generated code removal (see Section~\ref{sec:corpus}), to avoid situations where identical methods appear in both training and test sets.  Taken together, we expect that the nominal performance scores for all approaches will be far lower in our experiments.

Other related AI/data approaches in generating summaries of subroutines includes 1) work by Hu~\emph{et al.}~\cite{hu2018summarizing} focusing on creating summaries from sequences of API calls, and 2) CODE-NN by Iyer~\emph{et al.}~\cite{iyer2016summarizing} which, similar to \cite{hu2018deep}, creates a custom word representation of code which it then feeds to an off-the-shelf seq2seq model.

There is also related work outside the task of subroutine summaries.  Jiang~\emph{et al.}~\cite{jiang2017automatically}  and Loyola~\emph{et al.}~\cite{loyola2017neural} generate descriptions of code changes (i.e. commit messages).  Allamanis~\emph{et al.}~\cite{allamanis2016convolutional} predict a name for a subroutine from the body of a subroutine.  Oda~\emph{et al.}~\cite{oda2015learning} create pseudocode from source code by adapting statistical machine translation.  Yin~\emph{et al.}~\cite{yin2018mining}, Movshovitz~\emph{et al.}~\cite{movshovitz2013natural}, and Allamanis~\emph{et al.}~\cite{allamanis2015bimodal} target comments of short snippets of code, a task facilitated by public datasets~\cite{barone2017parallel}.  Gu~\emph{et al.}~\cite{gu2018deep} have demonstrated using a neural model for source code search, another task growing in popularity and facilitated by public datasets~\cite{yao2018staqc}.  Of note is that the attentional encoder-decoder seq2seq model originally described by Bahdanau~\emph{et al.}~\cite{bahdanau2014neural} is at the core of many of these papers, as it provides strong baseline performance even for many software engineering tasks.

\subsubsection{Heuristic/Template-based} Haiduc~\emph{et al.}~\cite{haiduc2010supporting, haiduc2010use} is often cited as the first attempt to create text summaries of code, and indeed is the first to introduce the term ``source code summarization.''  These early approaches create extractive summaries by calculating the top-\emph{n} keywords with metrics such as TF/IDF.  Shortly thereafter, work by Sridhara~\emph{et al.}~\cite{sridhara2010towards, sridhara2011automatically} adapted SWUM~\cite{hill2009automatically} (a technique for finding parts of speech of words in code) to create short summary phrases for source code using templates.  Another template-based solution by McBurney~\emph{et al.}~\cite{mcburney2016automatic} also used SWUM, but summarized a subroutine's context (defined as the functions that call or are called by a method) in addition to the method context.  Rodeghero~\emph{et al.}~\cite{rodeghero2015eye} made further improvements to content extraction for heuristic and template solutions by modifying the heuristics to mimic how human programmers read code with their eyes.  As in other research areas related to natural language generation~\cite{sutskever2011generating}, data-driven techniques have largely supplanted template-based techniques due to a much higher degree of flexibility and reduced human effort in template creation.  %McBurney~\emph{et al.}~\cite{mcburney2016automatic} demonstrated that template-based approaches have benefits but are far inferior to human-written summaries.%  A majority of work in code summarization since 2014 has focused on data-driven techniques.
We direct readers to a comprehensive survey by Nazar~\emph{et al.}~\cite{nazar2016summarizing}.% for template-based work on software artifacts other than subroutines.

\subsection{Neural Machine Translation}

The workhorse of most Neural Machine Translation (NMT) systems is the attentional encoder-decoder architecture~\cite{luong2015effective}.  This architecture originated in work by Bahdanau~\emph{et al.}~\cite{bahdanau2014neural} and is explained in great detail by a plethora of very highly-regarded sources~\cite{sutskever2014sequence, lecun2015deep, goodfellow2016deep, vaswani2017attention, johnson2017google}.  In this section, we cover only the concepts necessary to understand our approach at a high level.

In an encoder-decoder architecture, there are a minimum of two recurrent neural networks (RNNs).  The first, called the encoder, converts an arbitrary-length sequence into a single vector representation of a specified length.  The second, called the decoder, converts the vector representation given by the encoder into another arbitrary-length sequence.  The sequence inputted to the encoder is one language e.g. English, and the sequence from the decoder is another language e.g. French.

Encoder-decoder architectures learn to predict sentences one word at a time -- the decoder generally does not try to predict a whole sentence at once.  The way this usually works is that during training, instead of sending the network:

\vspace{-0.1cm}
{\small
\begin{verbatim}
[ cat on the table ] => [ chat sur la table ]
\end{verbatim}
}
\vspace{-0.1cm}

The network receives 1) the whole input sequence, 2) a sequence of output words so far, plus 3) the correct next word:

\vspace{-0.1cm}
{\small
\begin{verbatim}
[ cat on the table ]
   => [ chat 0 0 0 ] + [ sur ]
[ cat on the table ]
   => [ chat sur 0 0 ] + [ la ]
[ cat on the table ]
   => [ chat sur la 0 ] + [ table ]
\end{verbatim}
}
\vspace{-0.1cm}

During inference, the trained model is given an input sequence, which is used to predict the first word in the output sentence.  Then the input sentence is sent to the model again, along with the first prediction.  The decoder outputs a prediction for the second word in the sentence, and so on, until the decoder predicts an end-of-sentence token.  

The problem with this strategy is that the encoder is burdened with creating a vector representation suitable for prediction at every output step.  In reality, some words in the input sentence will be more important than others for a particular output.  E.g., `on' for `sur'.  This is the motivation for ``attentional'' encoder-decoder networks~\cite{bahdanau2014neural}.  Essentially what happens is that instead of a single vector representation of the input sentence, an attention mechanism is placed between the encoder and decoder.  That attention mechanism receives the encoder's state at every time step -- in the example above, four vectors for each of the four positions in the sentence.  The attention mechanism, in essence, selects which vector from the encoder to use, so that different decoder predictions receive input from different positions in the input sequence.  Our work builds on the attentional encoder-decoder strategy in key ways that we describe in the next section.

\section{Our Proposed Model}
\label{sec:model}

This section describes our proposed neural model.  The model assumes a typical NMT architecture in which the model is asked to predict one word at a time, as described in the previous section.

\subsection{Model Overview}

Our model is essentially an attentional encoder-decoder system, except with two encoders: one for code/text data and one for AST data.  In the spirit of maintaining simplicity where possible, we used embedding and recurrent layers of equal size for the encoders.  We concatenate the output of attention mechanisms for each encoder as depicted here:

\begin{figure}[!h]
	\centering
	\vspace{-0.35cm}
	\includegraphics[width=0.30\textwidth]{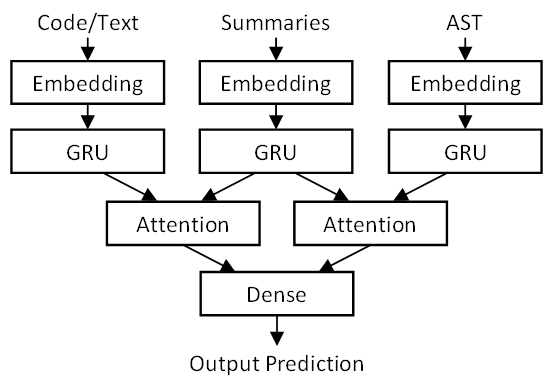}
	\vspace{-0.25cm}
	%\caption{An overview of this paper.}
	\label{fig:model_arch}
	%\vspace{-0.1cm}
\end{figure}

Precedent for combining different data sources comes heavily from image captioning~\cite{chen2015abc, yang2016stacked, johnson2016densecap, mao2014deep} (e.g. merging convolution image output with a list of tags).  One aim in this paper is to demonstrate how a similar concept is beneficial for code summarization, in contrast to the usual seq2seq application to SE data in which all information is put into one sequence.  We also hope to sow fertile ground for several areas of future work in creating unique processing techniques for each data type -- treating software's text and structure differently has a long tradition~\cite{marcus2003recovering}.

\subsection{Model Details}

To encourage reproducibility and for clarity, we explain our model as a walkthrough of our actual Keras implementation.  The following starts at line 29 in {\small\texttt{models/ast\_attendgru\_xtra.py}}; all code is available for download from our online appendix (Section~\ref{sec:reproducibility}).

\vspace{-0.1cm}
{\small \begin{verbatim}
txt_input = Input(shape=(self.txtlen,))
com_input = Input(shape=(self.comlen,))
ast_input = Input(shape=(self.astlen,))
\end{verbatim}}
\vspace{-0.1cm}

First, above, are three input layers corresponding to the code/text sequence, the comment sequence, and the flattened AST sequence.  We chose the sequence lengths as a balance between model size and coverage of the dataset.  The sequence sizes of 100 for code/text and AST, and 13 words for comment, each cover at least 80\% of the training set.  Shorter sequences are padded with zeros, and longer sequences are truncated.

\vspace{-0.1cm}
{\small \begin{verbatim}
ee = Embedding(output_dim=self.embdims, 
      input_dim=self.txtvocabsize)(txt_input)
se = Embedding(output_dim=self.embdims, 
      input_dim=self.astvocabsize)(ast_input)
\end{verbatim}}
\vspace{-0.1cm}

We start with a fairly common encoding structure, including embedding layers for each of our encoded input types (code/text and AST).  The embedding will output a shape of {\small \texttt{(batch\_size, txtvocabsize, embdims)}}.  What this means is that for every batch, each word in the sequence has one vector of length {\small \texttt{embdims}}.  For example, (200, 100, 100) means that for each of 200 examples in a batch, there are 100 words and each word is represented by a 100 length embedding vector.  We found two separate embeddings to have better performance than a unified embedding space.

\vspace{-0.1cm}
{\small \begin{verbatim}
ast_enc = CuDNNGRU(self.rnndims,
      return_state=True, return_sequences=False)
astout, sa = ast_enc(se)
\end{verbatim}}
\vspace{-0.1cm}

Next is a GRU layer with {\small \texttt{rnndims}} units (we found 256 to provide good results without oversizing the model) to serve as the AST encoding.  We used a CuDNNGRU to increase training speed, not for prediction performance.  The {\small \texttt{return\_state}} flag is necessary so that we get the final hidden state of the AST encoder.  The {\small \texttt{return\_sequences}} flag is necessary because we want the state \emph{at every cell} instead just the final state.  We need the state at every cell for the attention mechanism later.

\vspace{-0.1cm}
{\small \begin{verbatim}
txt_enc = CuDNNGRU(self.rnndims,
      return_state=True, return_sequences=True)
txtout, st = enc(ee, initial_state=sa)
\end{verbatim}}
\vspace{-0.1cm}

The code/text encoder operates in nearly the same way as the AST encoder, except that we start the code/text GRU with the final state of the AST GRU.  The effect is similar to if we had simply concatenated the inputs, except that 1) we keep separate embedding spaces, 2) we allow for attention to focus on each input differently rather than across input types, 3) we ensure that one input is not truncated by an excessively long sequence of the other input type, and 4) we ``keep the door open'' for further processing e.g. via convolution layers that would benefit one input type but not the other.  As we show in our evaluation, this is an important point for future work.

Tensor {\small \texttt{txtout}} would normally have shape {\small \texttt{(batch\_size, rnndims)}}, an {\small \texttt{rnndims}}-length vector representation of every input in the batch.  However, since we have {\small \texttt{return\_sequences}} enabled, {\small \texttt{encout}} has the shape {\small \texttt{(batch\_size, datvocabsize, rnndims)}}, which is the {\small \texttt{rnndims}}-length vector at every time-step.  That is, the {\small \texttt{rnndims}}-length vector at every word in the sequence.
So we see the status of the output vector as it changes with each word in the sequence.
We also have {\small \texttt{return\_state}} enabled, which just means that we get {\small \texttt{st}}, the {\small \texttt{rnndims}} vector from the last cell.  This is a GRU, so this {\small \texttt{st}} is the same as the output vector, but we get it here anyway for convenience, to use as the initial state in the decoder.

\vspace{-0.1cm}
{\small \begin{verbatim}
de = Embedding(output_dim=self.embdims, 
      input_dim=self.comvocabsize)(com_input)
dec = CuDNNGRU(self.rnndims,
      return_sequences=True)
decout = dec(de, initial_state=st)
\end{verbatim}}
\vspace{-0.2cm}

The decoder is as described in many papers on NMT: a dedicated embedding space followed by a recurrent layer.  We start the decoder with the final state of the code/text RNN.

\vspace{-0.2cm}
{\small \begin{verbatim}
txt_attn = dot([decout, txtout], axes=[2, 2])
txt_attn = Activation('softmax')(txt_attn)
\end{verbatim}}
\vspace{-0.2cm}

The next step is the code/text attention mechanism, with a design similar to that described by Luong~\emph{et al.}~\cite{luong2015effective}.  First, we take the dot product of the decoder and code/text encoder output.  The output shape of {\small \texttt{decout}} is, e.g., {\small \texttt{(batch\_size, 13, 256)}} and {\small \texttt{txtout}} is {\small \texttt{(batch\_size, 100, 256)}}.

The axis 2 of {\small \texttt{decout}} is 256 long.  The axis 2 of {\small \texttt{txtout}} is also 256 long.  So by computing the dot product along axis 2 in both, we get a tensor of shape {\small \texttt{(batch\_size, 13, 100)}}.  For one example in the batch, we get {\small \texttt{decout}} of (13, 256) and {\small \texttt{txtout}} (256, 100).

\vspace{-0.2cm}
\begin{table}[h!]
	\begin{tabular}{p{2.1cm}lp{2.1cm}lp{2.1cm}}
	
	~~~~decout (axis 2) & &  ~~~~txtout (axis 2) & & ~~~~~~txt\_attn \\
	
	\begin{tabular}{p{0.00cm}p{0.00cm}p{0.00cm}p{0.00cm}p{0.00cm}}
		   & 1        & 2        & ..        & 256   \\
		1  & \multicolumn{4}{l}{$v1 \longrightarrow$ } \\
		2  & \multicolumn{4}{l}{$v2 \longrightarrow$ } \\
		.. &          &          &           &       \\
		13 &          &          &           &           
	\end{tabular}
      & * &
	\begin{tabular}{p{0.00cm}p{0.00cm}p{0.00cm}p{0.00cm}p{0.00cm}}
		    & 1        & 2        & ..        & 100   \\
		1   & $v3$                            & $v4$  &    &     \\
		2   & $\downarrow$ & $\downarrow$ &   &     \\
		..  &          &          &           &       \\
		256 &          &          &           &           
    \end{tabular}
      & = &
	\begin{tabular}{p{0.00cm}p{0.00cm}p{0.00cm}p{0.00cm}p{0.00cm}}
		   & 1        & 2        & ..        & 100   \\
		1  & a		  & b		 & 			 & \\
		2  & c		  & d		 & 			 & \\
		.. &          &          &           &       \\
		13 &          &          &           &           
	\end{tabular}
      \\

    \end{tabular}
\end{table}
\vspace{-0.3cm}

Where {\small \texttt{a}} is the dot product of vectors {\small \texttt{v1}} and {\small \texttt{v3}}, and {\small \texttt{b}} is the dot product of {\small \texttt{v1}} and {\small \texttt{v4}}, etc.  

The result is that each of the 13 positions in the decoder sequence is now
represented by a 100-length vector.  Each value in the 100-length vector reflects the 
similarity between the element in the decoder sequence and the element in the encoder
sequence.  I.e. {\small \texttt{b}} above reflects how similar element 1 in the decoder sequence
is similar to element 2 in the code/text encoder sequence.  The 100-length vector for each of the 13 input positions reflects how much that a given input position is similar (should ``pay attention to'') a position in
the output.

Then we apply a softmax to each of the 13 (100-length) vectors.  The effect is
to exaggerate the ``most similar'' things, so that ``more attention'' will be paid to the 
more-similar input vectors -- the network learns during training to make them more similar.  Note that the dot product here is not normalized, so it is not necessarily equivalent to
cosine similarity.

%To be clear, the output shape of {\small \texttt{txt\_attn}} is {\small \texttt{(batch\_size, comvocabsize, txtvocabsize)}}.

\vspace{-0.2cm}
{\small \begin{verbatim}
txt_context=dot([txt_attn, txtout],axes=[2, 1])
\end{verbatim}}
\vspace{-0.2cm}

Next, we make use of the attention vectors by using them to create the context vectors for the code/text input.  To do that, we scale the encoder vectors by the attention vectors.  This is how we ``pay 
attention'' to particular areas of input for specific outputs.  The above line of code
takes {\small \texttt{txt\_attn}}, with shape {\small \texttt{(batch\_size, 13, 100)}}, and computes the dot product with
txtout {\small \texttt{(batch\_size, 100, 256)}}.  Recall that the encoder has {\small \texttt{txtvocabsize}}; 100 elements since it takes a sequence of 100 words.  Axis 1 of this tensor means ``for each element of the input sequence.''

The multiplication, for each sample in the batch, is:

\vspace{-0.2cm}
\begin{table}[h!]
	\begin{tabular}{p{2.1cm}lp{2.1cm}lp{2.1cm}}
		
		~~~~txt\_attn (axis 2) & &  ~~~~txtout (axis 1) & & ~~~~~txt\_context \\
		
		\begin{tabular}{p{0.00cm}p{0.00cm}p{0.00cm}p{0.00cm}p{0.00cm}}
			& 1        & 2        & ..        & 100   \\
			1  & \multicolumn{4}{l}{$v1 \longrightarrow$ } \\
			2  & \multicolumn{4}{l}{$v2 \longrightarrow$ } \\
			.. &          &          &           &       \\
			13 &          &          &           &           
		\end{tabular}
		& * &
	\begin{tabular}{p{0.00cm}p{0.00cm}p{0.00cm}p{0.00cm}p{0.00cm}}
	& 1        & 2        & ..        & 256   \\
	1   & $v3$                            & $v4$  &    &     \\
	2   & $\downarrow$ & $\downarrow$ &    &     \\
	..  &          &          &           &       \\
	100 &          &          &           &           
\end{tabular}
		& = &
		\begin{tabular}{p{0.00cm}p{0.00cm}p{0.00cm}p{0.00cm}p{0.00cm}}
			   & 1        & 2        & ..        & 256   \\
			1  & a		  & b		 & 			 & \\
			2  & c		  & d		 & 			 & \\
			.. &          &          &           &       \\
			13 &          &          &           &           
		\end{tabular}
		\\
		
	\end{tabular}
\end{table}
\vspace{-0.3cm}

The result is a context \emph{matrix} that has one context vector for each element in
the output sequence.  This is different than the vanilla sequence to sequence
approach, which has only one context vector used for every output.  Each output sequence location has its own context vector.  This vector is created from the most attended-to part of the encoder sequence.

\vspace{-0.1cm}
{\small \begin{verbatim}
ast_attn = dot([astout, encout], axes=[2, 2])
ast_attn = Activation('softmax')(ast_attn)
ast_context =
      dot([ast_attn, txtout], axes=[2, 1])
\end{verbatim}}
\vspace{-0.1cm}

We perform the same attention operations to the AST encoding as we do for the code/text encoding.

\vspace{-0.1cm}
{\small \begin{verbatim}
context = concatenate(
          [txt_context, ast_context, decout])
\end{verbatim}}
\vspace{-0.1cm}

But, we still need to combine the code/text and AST context with the decoder sequence information.  This is important because we send each word one at a time, as noted in the previous section.  The model gets to look at the previous words in the sentence in addition to the words in the encoder sequences.  It does not have the burden of predicting the entire output sequence all at once.  Technically, what we have here are two context matrices with shape {\small \texttt{(batch\_size, 13, 256)}} and a {\small \texttt{decout}} with shape {\small \texttt{(batch\_size, 13, 256)}}.  The default axis is -1, which means the last part of the shape (the 256 one in this case).  This creates a tensor of shape  {\small \texttt{(batch\_size, 13, 768)}}: one 768-length vector for each of the 13 input elements instead of three 256-length vectors.

\vspace{-0.1cm}
{\small \begin{verbatim}
out = TimeDistributed(Dense(self.rnndims,
      activation="relu"))(context)
\end{verbatim}}
\vspace{-0.1cm}

We are nearing the point of predicting a next word.  A TimeDistributed layer provides one dense layer per vector in the context matrix.  The result is one {\small \texttt{rnndims}}-length vector for every element in the decoder sequence.  For example, one 256-length vector for each of the 13 positions in the decoder sequence.  Essentially, this creates one predictor for each of the 13 decoder positions.

\vspace{-0.1cm}
{\small \begin{verbatim}
out = Flatten()(out)
out = Dense(self.comvocabsize, 
      activation="softmax")(out)
\end{verbatim}}
\vspace{-0.1cm}

However, we are trying to output a single word, the next word in the sequence.  Ultimately we need a single output vector of length {\small \texttt{comsvocabsize}}.  So we first flatten the (13, 256) matrix into a single (3328) vector, then we use a dense output layer of length {\small \texttt{comsvocabsize}}, and apply softmax.

\vspace{-0.1cm}
{\small \begin{verbatim}
model = Model(inputs=[txt_input, com_input, 
                      ast_input], outputs=out)
\end{verbatim}}
\vspace{-0.1cm}

The result is a model with code/text, AST, and comment sequence inputs, and a predicted next word in the comment sequence as output.
\vspace{-0.2cm}
\subsection{Hardware Details}

The hardware on which we implemented, trained, and tested our model included one Xeon E5-1650v4 CPU, 64gb RAM, and two Quadro P5000 GPUs.  It was necessary to train on GPUs with 16gb VRAM due to the large size of our model.

%At this point we are on page four.  The rest of the paper will go something like this:
%
%\begin{itemize}
%
%	\item Proposed Model: Two pages of so describing the model in detail.  I imagine something along the lines of detail in the attendgru comments I wrote a while back.  We can go through every line in our model implementation and explain the rationale and mechanism of action.  End near start of page six.
%	
%	\item Corpus Preparation: Half page of description of our dataset.  End near middle of page six.
%	
%	\item Evaluation: 1.5 pages about our research questions, methodology, etc.  Split over two experiments: standard and challenge.  End at end of page seven.
%	
%	\item Evaluation results: 1 page about our results, BLEU scores, etc.  End at end of page eight.
%	
%	\item Discussion: 1.5 pages of examples and answering why we get the results we do.  End around middle of page ten.
%	
%	\item Conclusion/Reproducibility/Acknowledgements: Half page of those.
%	
%\end{itemize}
%

%\vspace{-0.1cm}
\section{Corpus Preparation}
\label{sec:corpus}

We prepared a large corpus of Java methods from the Sourcerer repository provided by Lopes~\emph{et al.}~\cite{Lopes+Bajracharya+Ossher+Baldi:2010}.  The repository contains over 51 million Java methods from over 50000 projects.  We considered updating the repository with new downloads from GitHub, but we found that the Sourcerer dataset was quite thorough, leading to a large amount of overlap with newer projects that could not be eliminated (due to name changes, code cloning, etc.).  This overlap could lead to major validity problems for our experiments (e.g., if testing samples were inadvertently placed in the training set).  We decided to use the Sourcerer projects exclusively.

Significant preparation was necessary to make the repository a suitable dataset for applications of NMT, and we view this preparation as an important contribution of this paper to the research field (unlike in the NMT research area, there are relatively few curated datasets for code summarization).  After downloading the archives, we used a toolkit provided by McMillan~\emph{et al.}~\cite{mcmillan2011portfolio} to extract the Java methods and organize them into a SQL database.  Then we filtered for methods that were preceded by JavaDoc comments (indicated by {\small $/**$}).  We used only comments intended as JavaDocs, because there is an assumption that the first sentence in the comment will be a summary of the method's behavior~\cite{kramer1999api}.  Then we extracted the first sentence by looking for the first period, or the first newline if no period was present.  Next we used the {\small \texttt{langdetect}} library to remove comments not in English.  About 4m methods remained after these steps.

A potential problem was auto-generated code.  Auto-generated code is a problem because both the code and comments created by auto-generators tend to be very similar.  If nearly-identical code is in the training and testing sets, the model will learn these cases easily, which could simultaneously reduce performance on the ``real'' examples while falsely inflating performance metrics such as BLEU, since the metrics would reward the model for correctly identifying the duplicate cases.  Happily, the solution is fairly simple: we remove any methods from files that include phrases such as ``generated by'' suggested by Shimonaka~\emph{et al.}~\cite{shimonaka2016identifying}.  This filter is quite aggressive, as it reduced the dataset size to around 2m methods, and on manual inspection we found no cases of auto-generated code.  In fact, a majority of the filtered methods were exact duplicates (around 100k unique examples out of {\small$\sim$}2m removed methods).  But because comments to auto-generated code are often still meaningful, we added one copy of each of the 100k unique examples back into the dataset, and ensured that they were in the training set only (so we did not attempt to test against auto-generated comments).  The result is a dataset of around 2.1m methods.

Our other preprocessing steps followed the practice of many software engineering papers.  We split the code and comments on camel case and underscore, removed non-alpha characters, and set to lower case.  We did not perform stemming.

We then split the dataset \emph{by project} into training, validation, and test sets.  By ``by project'' we mean that we randomly divided the projects into the three groups: 90\% of projects into training, 5\% into validation, and 5\% into testing.  Then all the methods from a project went into the group assigned to its project.  A side effect is that since projects have different numbers of methods, 91\% of methods are in training, 4.8\% in validation, and 4.2\% in testing.  But this slight variation is necessary to maintain a realistic situation.  As mentioned in Section~\ref{sec:background}, we respectfully believe that not splitting by project and not removing auto-generated code are mistakes made by a vast majority of previous NMT applications to code summarization, and artificially inflates the reported scores (for example, SBT is reported to have 38 BLEU, versus 14 BLEU with the same technique in our evaluation).

To obtain the ASTs, we first used {\small \texttt{srcml}}~\cite{collard2011lightweight} to extract an XML representation of each method.  Then we built a tool to convert the XML representation into the flattened SBT representation, to generate SBT-formatted output described by Hu~\emph{et al.}~\cite{hu2018deep}.  Finally, we created our own modification of SBT in which all the code structure remained intact, but in which we replaced all words (except official Java API class names) in the code to a special $<$OTHER$>$ token.  We call this SBT-AO for SBT AST only.  We use this modification to simulate the case when only an AST can be extracted.

From this corpus of Java methods, we create two datasets:

\begin{itemize}
	\item The \textbf{standard dataset} contains three elements for each Java method: 1) the pre-processed Java source code for the method, 2) the pre-processed comment, and 3) the SBT-AO representation of the Java code.

	\vspace{0.1cm}

	\item The \textbf{challenge dataset} contains two elements for each method: 1) the pre-processed comment, and 2) the SBT-AO representation of the Java code.
\end{itemize}

Technically, we also have a third dataset containing the default SBT representation (with code words) and the pre-processed comment, which we use for experiments to compare our approach to the baselines.  However, the standard and challenge datasets are our focus in this paper, intended to compare the case when internal documentation is available, and the much more difficult case with only an AST.
\section{Evaluation}
\label{sec:eval}

This section covers our evaluation, comparing our approach to baselines over the standard and challenge datasets.

\subsection{Research Questions}

Our research objective is to determine the performance difference between our approach and competitive baseline approaches in two situations that we explore through these Research Questions (RQs):

\begin{description}
	\item[RQ$_{1}$] What is the difference in performance between our approach and competitive approaches in the ``standard'' situation, assuming internal documentation?
	
	\vspace{0.05cm}
	
	\item[RQ$_{2}$] What is performance of our approach in the ``challenge'' situation, assuming an AST only?
\end{description}

The rationale for these RQs was largely covered in the Introduction and Background sections.  Essentially, existing applications of NMT for the problem of code summarization almost entirely rely on the programmer writing meaningful internal documentation such as identifier names.  As we will show, this assumption makes the problem ``easy'' for seq2seq NMT models, since many methods have internal documentation that is very similar to the summary comment (a phenomenon also observed by Tan~\emph{et al.}~\cite{6200082} and Louis~\emph{et al.}~\cite{Louis2018DeepLT}).  We ask RQ$_1$ in order to study the performance of our approach under this assumption.

In contrast, we ask RQ$_2$ because the assumption of internal documentation is often not valid.  Very often, only the bytecode is available, or programmers neglect to write good internal documentation, or code has even been obfuscated deliberately.  In these cases, it is usually still possible to extract an AST for a method, even if it contains no meaningful words.  In principle, the structure of a program is all that is necessary to understand it, since ultimately that is what defines the behavior of the program.  In practice, it is very difficult to connect structure directly to high-level concepts described in summaries.  We seek to quantify a baseline performance level with our approach (since, to our knowledge, no published approach functions in this situation).

\subsection{Baselines}

To answer RQ$_1$ (the standard experiment), we compare our approach to three baselines.  One baseline (which we call {\small \textbf{\texttt{attendgru}}}) is a generic attentional encoder-decoder model, to represent an application of a strong off-the-shelf approach from the NLP research area.  Note that there are a huge variety of NMT systems described in the NLP literature, but that a vast majority have an attentional encoder-decoder model at their heart (see Section~\ref{sec:background}).  To maintain an ``apples to apples'' comparison, the baseline is identical to the ``code/text'' encoder in our approach (the decoder is identical as well).  In essence, the baseline is the same as our proposed approach, except without the AST encoder and associated concatenation.  While we could have chosen any number of approaches from NLP literature, it is very difficult to say up front which will perform best for code summarization, and we needed to ensure minimal differences to maximize validity of our results.  If, for example, we had used an architecture with an LSTM instead of a GRU in the encoder, we would have no way of knowing if the difference between our approach and the baseline were due to the AST information we added, or due to using an LSTM instead of a GRU.

A second baseline is the {\small \textbf{\texttt{SBT}}} approach presented by Hu~\emph{et al.}~\cite{hu2018deep}.  This approach was presented at ICPC'18, and (at the time of writing) represents the latest publication about source code summarization in a software engineering venue.  That paper used an LSTM-based encoder-decoder architecture based on a popular guide for building seq2seq NMT systems, but used their SBT representation of code instead of the source code only.  For our baseline, we use their SBT representation, but use the same GRU-based encoder-decoder from our NLP baseline, also to ensure an ``apples to apples'' comparison.  Since the model architecture is the same, we can safely attribute performance differences to the input format (e.g., SBT vs. code-only).

A third baseline is {\small \textbf{\texttt{codenn}}}, presented by Iyer~\emph{et al.}~\cite{iyer2016summarizing}.  Given the complexity of the approach, we used their publicly-available implementation.  The original paper describes only applications to SQL and C\#, but we noticed that their C\# parser extracted common code features that are also available in Java.  We made small modifications to the C\# parser so that it would function equivalently for Java.

We call our approach {\small \textbf{\texttt{ast-attendgru}}} in our experiments.  We used a greedy search algorithm for inference for all approaches, rather than beam search, to minimize the number of experimental variables and computation cost.
%\vspace{-0.1cm}
\subsection{Methodology}
%\vspace{-0.1cm}
Our methodology to answer both RQs is identical, and follows best practice established throughout the literature on NMT (see Section~\ref{sec:background}): for RQ$_1$, we train our approach and each baseline with the training set from the standard dataset for a total of 10 epochs.  Then, for each approach, we computed performance metrics for the model after each epoch against the validation set.  (In all cases, validation performance began to degrade after five or six epochs.)  Next we chose the model after the epoch with the highest validation performance, and computed performance metrics for this model against the testing set.  These testing results are the results we report in this paper.  Our methodology to answer RQ$_2$ differs only in that we trained and tested using the challenge dataset.

We report the performance metric BLEU~\cite{papineni2002bleu}, also in keeping with standard practice in NMT.  BLEU is a measure of the text similarity between predicted summaries and reference summaries.  We report a composite BLEU score in addition to BLEU$_1$ through BLEU$_4$ (BLEU$_n$ is a measure of the similarity of $n$-length subsequences, versus entire summary sentences).  Technically speaking, we used {\small \texttt{nltk.translate.bleu\_score}}~\cite{NLTKWebsite} in our implementation.

\subsection{Threats to Validity}

The primary threats to validity to this evaluation include: 1) Our dataset.  We use a very large dataset with millions of Java methods in order to maximize the generalizability of our results, but the possibility remains that we would obtain different results with a different dataset.  And, 2) we did not perform cross-validation.  We attempt to mitigate this risk by using random samples to split the training/validation/testing sets, a different split could result in different performance.  This risk is common among NMT experiments due to very high training computation costs (4+ hours per epoch).

\section{Evaluation Results}
\label{sec:results}

This section discusses our evaluation results and observations.  After answering our research questions, we explore examples to give an insight into how the network functions and why it works.  Note that we use these observations to build an ensemble method at the end of this paper.

\setcounter{figure}{0}
\begin{figure}[b!]
	\centering
	\vspace{-0.6cm}
	\includegraphics[width=0.35\textwidth]{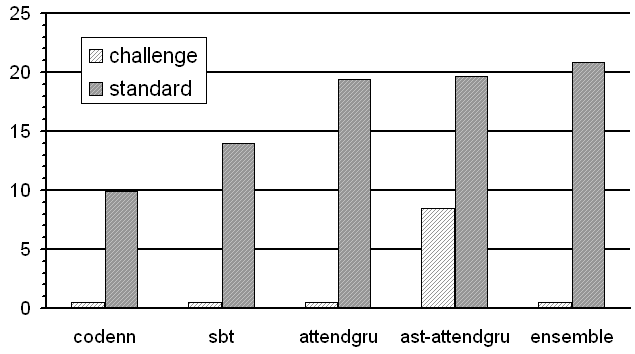}
	\vspace{0.0cm}
	%\caption{A}

	{\small
	\begin{tabular}{p{2cm}|p{0.5cm}p{0.5cm}p{0.5cm}p{0.5cm}p{0.5cm}|p{1cm}}
	model         & B & B1 & B2 & B3 & B4  & dataset                   \\ \hline
	ast-attendgru & 19.6 & 39.3 & 22.2 & 14.9 & 11.4 & \multirow{4}{*}{standard} \\
	attendgru     & 19.4  & 39.0  & 22.0  & 14.8  & 11.3  &                           \\
	sbt           & 14.0  & 31.8  & 16.0  & 10.1  & 7.5  &                           \\
	codenn        & 9.95  & 21.2  & 9.7  & 7.6  & 6.3  &                           \\ \hline
	ast-attendgru & 9.47  & 25.7  & 11.0  & 6.1  & 4.7  & challenge                
	\end{tabular}
	}
	
	\caption{{\small Below are BLEU1-4 scores and the composite BLEU score for each approach and dataset.  Above, the chart depicts the composite scores only.  We observe that attendgru and ast-attendgru perform equally in terms of BLEU score on the standard set, though we improve it with an ensemble decoder in Section~\ref{sec:ensemble}.}}
	\label{fig:resultsoverview}
	
	\vspace{-0.5cm}
\end{figure}
%\vspace{-0.1cm}
\subsection{RQ$_1$: Standard Experiment}

We found in the standard experiment that {\small \texttt{ast-attendgru}} and {\small \texttt{attendgru}} obtain roughly equal performance in terms of BLEU score, but provide orthogonal results, as we will explain in this section and the example in subsection \emph{C}.

In terms of BLEU score, {\small \texttt{ast-attendgru}} and {\small \texttt{attendgru}} are roughly equal in performance: 19.6 BLEU vs 19.4 BLEU.  {\small \texttt{SBT}} is lower, at about 14 BLEU, and {\small \texttt{codenn}} is about 10 BLEU.  Figure~\ref{fig:resultsoverview} includes a table with the full BLEU results for each result (and additional data in our online appendix).  For {\small \texttt{SBT}}, the results conflicted with our expectations based on the presenting paper~\cite{hu2018deep}, in which {\small \texttt{SBT}} outperformed a standard seq2seq model like {\small \texttt{attendgru}}.  We see two possible explanations: First, even though our seq2seq baseline implementation represents a standard approach, there are a few architectural differences from the paper by Hu~\emph{et al.}~\cite{hu2018deep}, such as different embedding vector sizes.  While we did not observe major changes in the results from these architectural differences in our own pilot studies, it is possible that ``one's mileage may vary'' depending on the dataset.  Second, as we note in Sections~\ref{sec:background} and~\ref{sec:corpus}, the previous study did not split by project, so methods in the same project will be in the training and test set.  The very high reported BLEU scores in~\cite{hu2018deep} could be explained by overloaded methods with very similar structure -- {\small \texttt{SBT}} would detect a function in the test set with a very similar AST to an overloaded method in the same project in the training set.

The improvement by all approaches over {\small \texttt{codenn}} matches expectations from previous experiments.  The {\small \texttt{codenn}} approach was intended as a versatile technique for both code search and summarization, and was a relatively early attempt at applying NMT to the code summarization problem.  In addition, it was designed for C\# and SQL datasets; we adapted it to Java as described in the previous section.

A key observation of the standard experiment is that {\small \texttt{ast-attendgru}} and {\small \texttt{attendgru}} provide \textbf{orthogonal} predictions -- there is a set of methods in which one performs better, and a different set in which the other has higher performance.  While {\small \texttt{ast-attendgru}} is slightly ahead of {\small \texttt{attendgru}}, we do not view a 0.2 BLEU difference a major improvement in and of itself.  Normally we would expect an approach to outperform a different approach by some margin across a majority of the examples (i.e. non-orthogonal performance), and this is indeed what we observe when comparing {\small \texttt{ast-attendgru}} to {\small \texttt{SBT}}, as shown on the left below (around 60k methods in which {\small \texttt{ast-attendgru}} performed better, vs. 20k for {\small \texttt{SBT}}):

\begin{figure}[h!]
	\vspace{-0.4cm}
	\centering
	\includegraphics[width=0.49\textwidth]{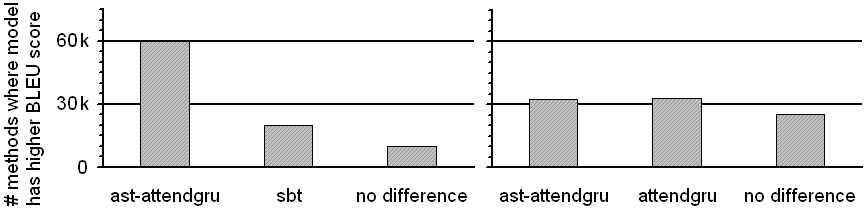}
	\vspace{-0.8cm}
\end{figure}

But what we observe for {\small \texttt{ast-attendgru}} and  {\small \texttt{attendgru}} is that there are two sets of roughly 33k methods in the 91k test set in which one or another approach has higher performance (above, right).  In other words, among the predictions in which there was a difference between the approaches, {\small \texttt{ast-attendgru}} and gives better predictions (in terms of BLEU score) for about half, while {\small \texttt{attendgru}} performs better on about half.  Orthogonal performance makes these two approaches a good candidate for ensemble prediction, which we further explain in subsection \emph{C} and Section~\ref{sec:ensemble}.

\setcounter{figure}{2}
\begin{figure*}[b!]
	\vspace{-0.5cm}
	\centering
	\includegraphics[width=\textwidth]{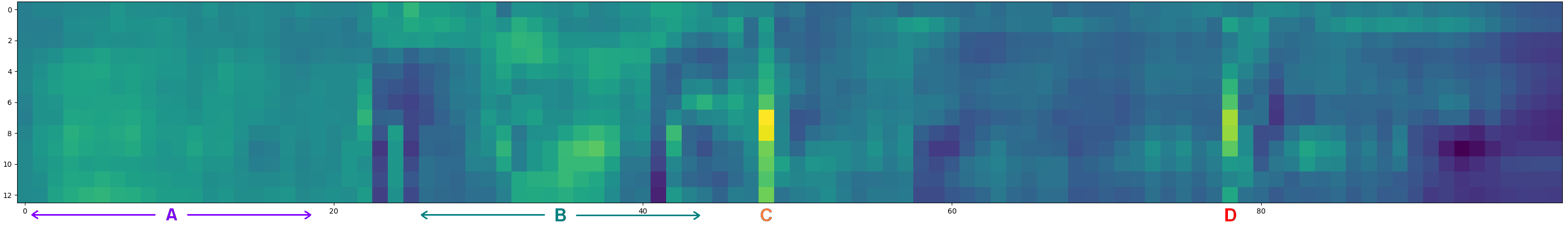}
	\vspace{-0.6cm}
	\caption{Heatmap of the attention layer in {\small \texttt{ast-attendgru}} for the AST input for Example 1.  The x-axis is the summary input and the y-axis is the AST (SBT-AO) input.  High activation (more yellow) indicates more attention paid to e.g. position 48 of the AST input.}
	\label{fig:astattend}
\end{figure*}
%\vspace{-0.1cm}
\subsection{RQ$_2$: Challenge Experiment}

We obtain a BLEU score of about 9.5 for {\small \texttt{ast-attendgru}} in the challenge experiment.  Note that the only difference between the standard and challenge experiments is that we trained and tested using the AST only, in the form of the SBT-AO representation fed to {\small \texttt{ast-attendgru}}.  Technically, there are other configurations that would produce the same result, such as using SBT-AO as input to {\small \texttt{attendgru}} instead of the source code.  Any of these configurations would meet our objective with this experiment of establishing performance for the scenario when only an AST is available.
%\vspace{-0.2cm}

\subsection{Explanation and Example}
%\vspace{-0.1cm}
Merely reporting BLEU scores leaves an open question as to what the scores mean in practice.  Consider these two examples from the standard and challenge experiments (method IDs align with our downloadable dataset for reproducibility).  We chose the following examples for illustrative purposes, and as an aid for explanation.  While relatively short, we feel that these methods provide a useful insight into how the models operate.  For a more in depth analysis, a human evaluation would be required, which is beyond the scope of this paper.

Example 1 is one of the cases where {\small \texttt{ast-attendgru}} succeeds when {\small \texttt{attendgru}} fails.  To understand why, recall that, in our model as with a majority of NMT systems, the system predicts a sentence one word at a time. For each word, the model receives information about the method (the code/text plus the AST for models that use it), along with each word that has been predicted so far.
\vspace{-0.2cm}
\begin{figure}[h!]
	%\centering
	\textbf{Example 1}, Method ID 49111725:
	%\vspace{-0.2cm}
	{\small
		
		\begin{verbatim}
public Config tokenUrl(String tokenUrl) {
    this.tokenUrl = tokenUrl;
    return this; }
		\end{verbatim}
		\vspace{-0.1cm}
		\begin{tabular}{llm{5.5cm}} %\hline
			\emph{reference}     &  & sets the token url \\ \hline
			ast-attendgru &  & sets the token url			\\ \cline{3-3}
			attendgru     & \multirow{1}{*}{stan.} & returns the url of the token   			\\ \cline{3-3}
			sbt           &  & sets the $<$UNK$>$    \\ \cline{1-3}
			ast-attendgru &  chal. & sets the value of the $<$UNK$>$ property       \\ \hline                           
		\end{tabular}
	
	\vspace{0.5cm}
	\emph{Tokenized code/text input:} $<$s$>$ public config token url string token url this token url token url return this $<$/s$>$

	\begingroup
	\linespread{1.5}%
	\selectfont
	\emph{SBT-AO input:} \textcolor{Fuchsia}{( unit ( function ( specifier ) specifier\_OTHER ( type ( name ) name\_OTHER ) type ( name ) name\_OTHER} \textcolor{Emerald}{( parameter\_list ( parameter ( decl ( type ( name ) name\_String ) type ( name ) name\_OTHER ) decl ) parameter ) parameter\_list} ( block ( \textcolor{Orange}{expr\_stmt} ( expr ( name ( name ) name\_OTHER ( operator ) operator\_OTHER ( name ) name\_OTHER ) name ( operator ) operator\_OTHER ( name ) name\_OTHER ) expr ) \textcolor{BrickRed}{expr\_stmt} ( return ( expr ( name ) name\_OTHER ) expr ) return ) block ) function ) unit
	\endgroup

\vspace{0.3cm}
\setcounter{figure}{1}

	\centering
	\begin{subfigure}[b]{0.40\linewidth}
	\includegraphics[width=\textwidth]{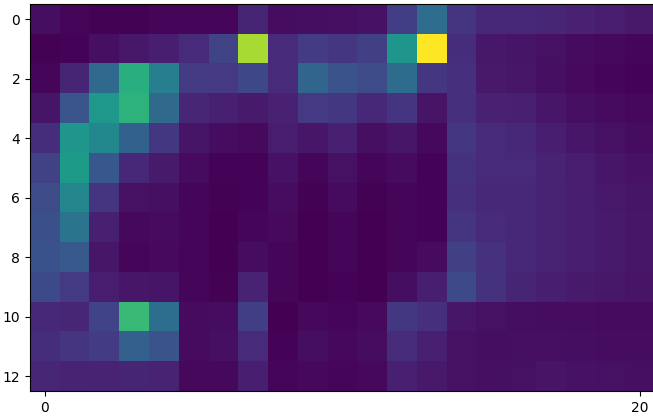}
	   
	~~~~~~~~~(a) {\small \texttt{attendgru}}
	\end{subfigure}
	~
	\begin{subfigure}[b]{0.20\textwidth}
	\includegraphics[width=\textwidth]{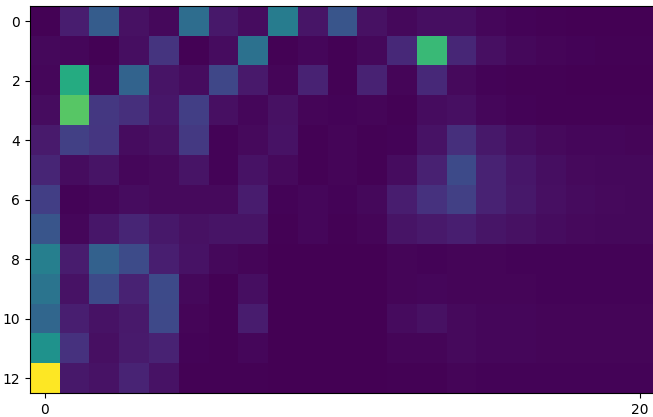}
	
	~~~~~~(b) {\small \texttt{ast-attendgru}}
	\end{subfigure}
    \vspace{-0.1cm}
	\caption{{\small Heatmaps of the attention layer in (a) {\small \texttt{attendgru}} and (b) {\small \texttt{ast-attendgru}} for the code/text input for Example 1.  The x-axis is the 13 positions in the summary input.  The y-axis is the 100 positions in the code input.  Images are truncated to code input length.}}
	\label{fig:codeattend}
	}
\end{figure}
\noindent
    So to predict ``token'', {\small \texttt{ast-attendgru}} would receive the code/text, the AST, and the phrase ``sets the''.In contrast, {\small \texttt{attendgru}} only receives the code/text and ``sets the''.  To predict the first word, ``sets'', {\small \texttt{attendgru}} only knows that it is the start of the sentence (indicated by a start-of-sentence $<$s$>$ token), and the code/text input.  To help make the prediction {\small \texttt{attendgru}} is equipped with an attention layer learned during training to attend to certain parts of the input.  That layer is depicted in Figure~\ref{fig:codeattend}(a).  Note that there is high activation (bright yellow) in position (14, 1), indicating significant attention paid to location 14 in the code/text input: this is the word return.  What has happened is that, during training, the model saw many examples of getter methods that were only a few lines and ended with a return.
  
  In many cases, the model could rely on very explicit method names, such as {\small \texttt{getPlayerScore}} (method ID 38221679).  {\small \texttt{attendgru}} performed remarkably well in these cases, as the situation is quite like natural language -- it learns to align words in the input vocabulary to words in the target vocabulary, and where they belong in a sentence.  However, in cases such as Example 1 where the method name does not clearly state what the method should do (the name {\small \texttt{tokenUrl}} is not obviously a setter), {\small \texttt{attendgru}} struggles to choose the right words, even if, as in Example 1, it correctly identifies the subject of the action (``url of the token'').
  \vspace{-0.05cm}
  
  These situations are where the AST is beneficial.  The code/text activation layer for {\small \texttt{ast-attendgru}} attends heavily to the start of sentence token (note column 0 in Figure~\ref{fig:codeattend}(b)), which, since $<$s$>$ is the start of every sentence, probably acts like a ``not sure'' signal.  But the model also has the AST input.  Figure~\ref{fig:astattend} shows the AST attention layer of {\small \texttt{ast-attendgru}} when trying to predict the first word.  There are four areas of interest that help elucidate how the model processes the structure of the model, denoted A through D in the figure, and color-coded to the corresponding areas in the AST input.  First, area A, is the portion of the method signature prior to the parameter.  Recall that our AST representation is structure only, so almost all methods will start the same way.  So as expected, the attention in area A is largely formless.  The heatmap shows much more definition in area B.  It is the parameter list, and the model has likely learned that short methods with parameter lists tend to be setters.  The model activates very heavily at locations C and D, which are the start and end of the expr\_stmt AST node.  A very common situation in the training set is that a short method with a parameter and an assignment is a setter.  The model has learned this and chose ``sets'' as the first word.

All of the models with AST input correctly chose ``sets''.  {\small \texttt{SBT}} found that the method is a setter, but could not determine what was being set -- we attribute this behavior to the fact that the SBT representation blends the code/text and structural information into a single input, which creates a challenge for the model to learn orthogonal types of information in the same vector space (which work in other areas e.g. image captioning implies is not advisable~\cite{vinyals2015show}).  While there is not space in this paper to explore fully, we note that even {\small \texttt{ast-attendgru}} during the challenge experiment correctly characterized the method as setting the value of a property, generating an unknown token when it could not determine which property.  In fact, {\small \texttt{ast-attendgru}} correctly predicted the first word of the summary (which is usually a verb) 33\% of the time during the challenge experiment, compared to 52\% of the time in the standard experiment.  Briefly consider Example 2:

\vspace{-0.3cm}
\begin{figure}[h!]
	%\centering
	%\vspace{-0.1cm}
	\textbf{Example 2}, Method ID 40490666:
	\vspace{-0.25cm}
	{\small
		
		\begin{verbatim}
public void disconnect() {
    try {
        socket.flush();
        socket.close();
        connected = false;
        notifyDisconnect();
    } catch (IOException ex) {
        ex.printStackTrace(); } }
		\end{verbatim}
		
		\begin{tabular}{llm{5.5cm}} %\hline
			\emph{reference}     &  & closes the socket for reconnection \\ \hline
			ast-attendgru &  & disconnect from the server			\\ \cline{3-3}
			attendgru     & \multirow{1}{*}{stan.} & disconnects from the server   			\\ \cline{3-3}
			sbt           &  & disconnect from the server    \\ \cline{1-3}
			ast-attendgru &  chal. & closes the connection       \\ \hline                           
		\end{tabular}
	}
	
\end{figure}
\vspace{-0.3cm}

All approaches performed well for this method, but for different reasons.  {\small \texttt{attendgru}} linked the method name to the verb ``disconnects''.  {\small \texttt{SBT}} relied more on later features such as the call to notifyDisconnect().  Most interestingly, {\small \texttt{ast-attendgru}} performed best in the challenge experiment.  In exploring this result, we found a few methods with a similar AST (IDs 146827, 22838818, 28418561, 5785101).  All of these had a few lines in a try block followed by a short catch block, and 2-3 method calls and assignments to null or false in the try.  These methods had summaries like ``close the communication with the gps device'', ``stops the timer'', and ``disconnect from the current client'' -- all these methods deal with close and cleanup behavior.  The model probably learned this during training, and chose similar words for the summary.

In answering RQ$_1$, we found that {\small \texttt{attendgru}} and {\small \texttt{ast-attendgru}} performed better on different sets of methods.  While we are hesitant to overinterpret single examples, the examples in this section are consistent with numerous others in the dataset (we provide a script for randomly sampling examples called rand\_samples\_preds.py in our online appendix for interested readers).  The examples are also consistent with the interpretation that the off-the-shelf NMT system ({\small \texttt{attendgru}}) performs quite well in cases where the summaries are clear from the method signature, and in these cases the AST may be superfluous.  But, the model benefits from the AST in cases when words in the code/text input are not sufficient or clear.

\section{Ensemble Decoding and Future Work}
\label{sec:ensemble}

As a hint toward future work, we test a combination of the {\small \texttt{attendgru}} and {\small \texttt{ast-attendgru}} models using ensemble decoding.  The combination itself is straightforward: we compute an element-wise mean of the output vector of each model (the same trained models used in our evaluation).  The training and test procedure does not change, except that during prediction, we use the maximum value of the combined output vector, rather than just one output vector from one model.  This is the same ensemble decoding procedure implemented by OpenNMT~\cite{OpenNMTWebsite}, and is one of the most common of several options described by literature on multi-source NMT~\cite{garmash2016ensemble}.

Since we are combining output vectors, the models ``work together'' during prediction of every word -- it is not just choosing one model or another for the whole sentence.  The idea is that one model may assign similar weights in the output vector to two or more words, in cases where it performs less well.  And another model that performs better in that situation may assign more weight to a single word.  In our system, the hope is that {\small \texttt{attendgru}} will contribute more when code/text words are clear, but {\small \texttt{ast-attendgru}} will contribute more when they are unclear.

The ensemble decoding procedure improves performance to 20.9 BLEU, from 19.6 for {\small \texttt{ast-attendgru}} and 19.4 for {\small \texttt{attendgru}}.  This is more than a full BLEU point improvement, which is quite significant for a relatively simple procedure.  This result points us to future work including more advanced ensemble decoding (e.g. predicting when to use one model or another), optimizations to the network (e.g. dropout, parameter tuning), and, critically, using different data processing techniques on each type of input.
%\vspace{-0.1cm}
\section{Conclusion}
\label{sec:conclusion}

We have presented a neural model for generating natural language descriptions of subroutines.  We implement our model and evaluate it over a large dataset of Java methods.  We demonstrate that our model {\small \texttt{ast-attendgru}}, in terms of BLEU score, outperforms baselines from SE literature and is slightly ahead of a strong off-the-shelf approach from NLP literature.  We also demonstrate that and ensemble of our approach and the off-the-shelf NLP approach outperforms all other tested configurations.  We provide a walkthrough example to provide insight into how the models work -- we conclude that the default NMT system works well in situations where good internal documentation is provided, but less well when it is not provided, and that {\small \texttt{ast-attendgru}} assists in these cases.  We demonstrate how {\small \texttt{ast-attendgru}} can produce coherent predictions even with zero internal documentation.

%\vspace{-0.1cm}
\section{Reproducibility}
\label{sec:reproducibility}

Our dataset, code, models, and results are available via:

\url{https://bit.ly/2MLSxFg}

%{\small \emph{(Site anonymized for blind review.)}}
%\vspace{-0.1cm}
\section*{Acknowledgment}

{\small This  work  is  supported  in  part  by  the  NSF  CCF-1452959, CCF-1717607, and CNS-1510329 grants. Any opinions, findings, and conclusions expressed herein are the authors and do not necessarily reflect those of the sponsors}

%\section*{References}

\bibliographystyle{IEEEtran}
\bibliography{main}

\end{document}